\documentclass{rmf-d}
\usepackage{nopageno,rmfbib,multicol,times,epsf,amsmath,amssymb,cite}
\usepackage[latin1]{inputenc}
\usepackage[]{caption2}
\usepackage{graphics}
\usepackage{graphicx}
\def\rmfcornisa{Research in Physics \hfill\rmf\ {\bf ??} (*?*) ???--???
\hfill MES? A\~NO?}

\def\rmfcintilla{{\it Rev.\ Mex.\ Fis.\/} {\bf ??} (*?*) (????) ???--???}
\clearpage \rmfcaptionstyle 
\begin{document}
\title{Physics with intact protons at the LHC: from the odderon discovery to the sensitivity to beyond standard model physics
\vspace{-6pt}}
\author{Christophe Royon }
\address{The University of Kansas, Lawrence, USA}
\maketitle

\recibido{15 May 2023}{DD MM 2022 \vspace{-12pt} }

\begin{abstract}
\vspace{1em} We describe the discovery of the colorless $C$-odd gluonic compound, the odderon, by the D0 and TOTEM Collaborations by comparing elastic differential cross sections measured in $pp$ and $p \bar{p}$ interactions at high energies. We also discuss the reach on quartic anomalous couplings and the sensitivity to axion like particle production by using the LHC as a $\gamma \gamma$ collider and detecting the intact protons at high luminosity.
   \vspace{1em}
\end{abstract}
\keys{ key1, key2  \vspace{-4pt}}
\pacs{   \bf{\textit{pac1, pac2 }}    \vspace{-4pt}}
\begin{multicols}{2}

\section{Medal of the Mexican Society of Physics, Division of Particles and Fields}

It was a great pleasure and honor to receive the 2022 medal of the Mexican Society of Physics, Division of Particles and Fields during the annual meeting of the Division for ``his leadership in the discovery of odd-gluon state odderon from elastic proton-proton and proton-antiproton collisions at TOTEM and D0 detectors, his contributions to QCD and physics beyond the Standard Model (SM), and for his support to the Mexican High Energy Physics community". I would like to thank all the people who supported me in Mexico, especially Javier Murillo, and all my PhD students originally from Mexico or who studied there (Cristian from Hermosillo, one of my students who finished 1.5 year ago and who also received the 2023 APS Mitsuyoshi Tanaka Dissertation Award in Experimental Particle Physics, Luis from Honduras after a master in Cinvestav, Saray from Zacatecas also after a master in Cinvestav, Moises from Venezuela with a co-led PhD in Hermosillo), as well my other students and post-docs at the University of Kansas (Florian, Mats, Tommaso, Justin, Cole, Zach, Federico, L\'eo, William, Gauthier, Maxime, Guillaume, Nicola, Georgios, Laurent, Tim, Alexander, Hussein).

This medal will reinforce collaborations with Mexico for instance in the CMS, ALICE
collaborations at CERN (Javier from Sonora, Antonio from UNAM, etc)  about medical applications for flash beam therapy
(Javier from Sonora, Arturo from Puebla, etc), and phenomenology (Martin from Puebla, Pablo from Cinvestav, etc).

In this paper, we will describe in turn the topics that were related directly to the medal, namely the odderon discovery and the beyond-standard model physics using the LHC as a $\gamma \gamma$ collider.

\section{The discovery of the odderon}

\subsection{Elastic interactions}

\subsection{$pp$ and $p \bar{p}$ elastic scattering}
\begin{figure*}
\centering
\includegraphics[width=0.6\textwidth]{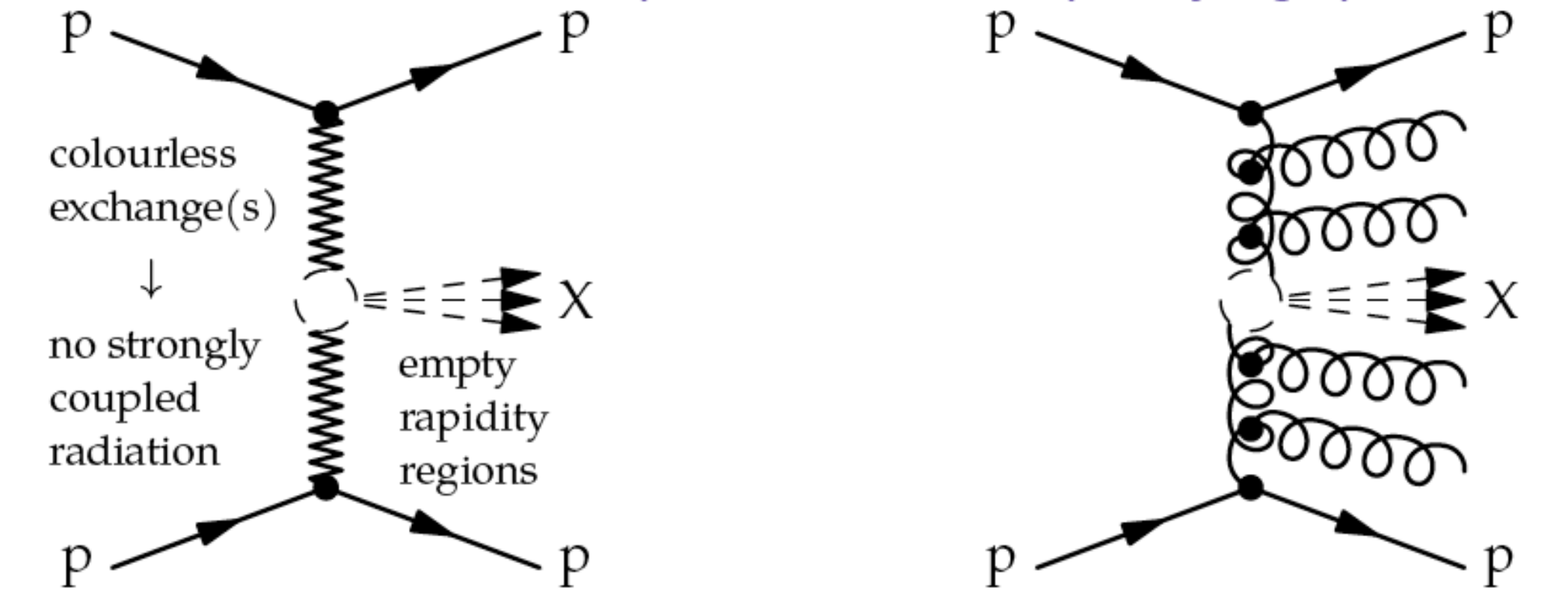}
\caption{Schematic of $pp$ elastic and inelastic interactions.}
\label{fig0}
\end{figure*}

Comparing elastic $pp$ and $p \bar{p}$ interactions is one of the best methods to look for the odderon~\cite{ourpaper}. Elastic interactions mean that protons and anti-protons remain intact after the collision. Fig.~\ref{fig0}, right, shows a typical $pp$ interaction with a gluon being exchanged between the two protons. There is some color exchange between the two protons and the exchanged gluon can radiate lots of gluons. It means that the protons cannot remain intact in the final state after interaction in that case, and they are destroyed. The case when protons can remain intact is depicted in Fig.~\ref{fig0}, right, and correspond to the exchange of 2, 3, 4, 5,... gluons but not one single gluon, since we need a colorless object to be exchanged. 

The colorless object can be either a pomeron or an odderon. Pomeron and odderon correspond respectively to positive and negative $C$ parity. The pomeron is made of an even number of gluons which leads to a ($+1$) parity whereas the odderon is made of an odd number of gluons corresponding to a ($-1$) parity~\cite{nicolescu,martynov,landshoff,leader}.
Scattering amplitudes for elastic $pp$ and $p \bar{p}$ interactions can be written as
\begin{eqnarray}
A_{pp} &=& Even~+~Odd  \\
A_{p \bar{p}} &=& Even~-~Odd .
\end{eqnarray}
and, from these equations, it is clear that observing a difference between $pp$ and $p \bar{p}$ interactions would be a clear way to observe the odderon. 

Some differences between $pp$ and $p\bar{p}$ elastic $d\sigma/dt$ cross sections of about 3$\sigma$ were already observed at ISR energies at CERN~\cite{ISR}. This was not considered to be a clean proof of the existence of the odderon since the situation is quite complicated at low energies where elastic interactions can be indeed explained via the exchange of pomeron and odderon, but also of additional reggeons and mesons. At TeV energies, the meson and reggeon exchanges become negligible and any difference between $pp$ and $p\bar{p}$ interactions would be a sign of the odderon.

The experimental signature of elastic interactions corresponding to the detection of the scattered intact protons and anti-protons after interaction in dedicated detectors called roman pots that can go close to the beam (down to 3$\sigma$). We use the LHC and Tevatron magnets as a spectrometer that deviates slightly the scattered protons and anti-protons from the beam since they lost part of their energy.

\begin{figure*}
\centering
\includegraphics[width=0.45\textwidth]{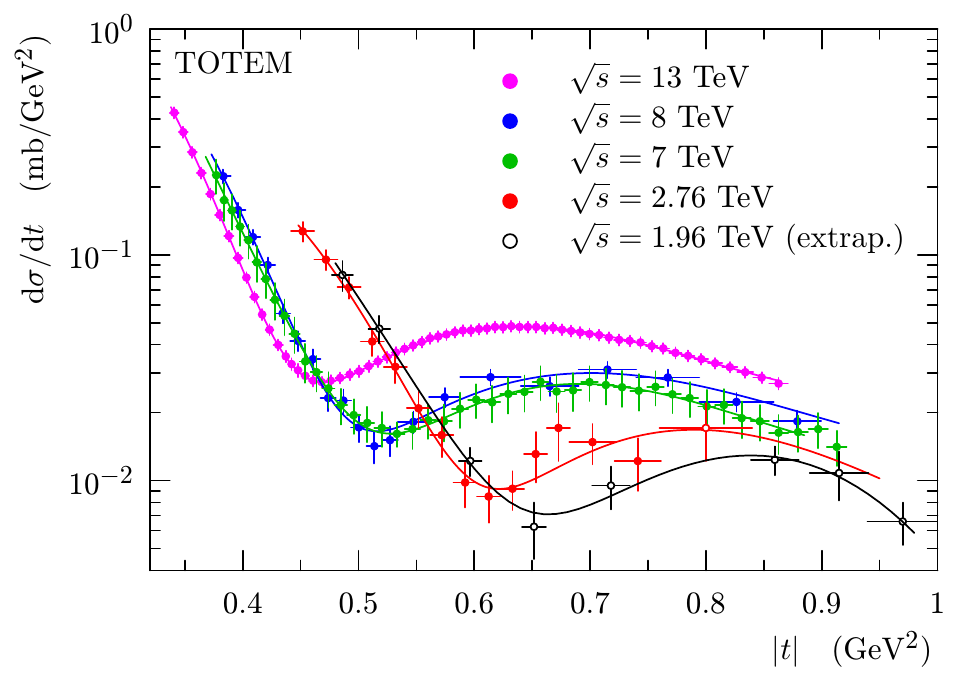}
\includegraphics[width=0.45\textwidth]{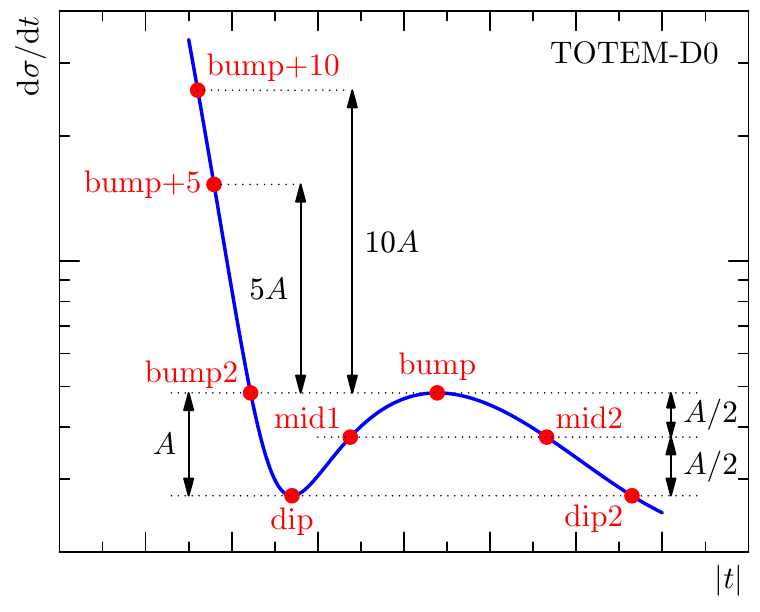}
\caption{Elastic $d \sigma/dt$ measurements from the TOTEM collaboration at 2.76, 7, 8 and 13 TeV at the LHC  extrapolated down to 1.96 TeV, the Tevatron center-of-mass energy. Characteristic points for elastic $pp$ $d\sigma/dt$.}
\label{fig1}
\end{figure*}

\subsection{D0 and TOTEM data and strategy to look for the odderon}

The D0 collaboration measured the elastic $p \bar{p}$ $d\sigma/dt$ cross section at the  1.96 TeV Tevatron for $0.26<|t|<1.2$ GeV$^2$~\cite{d0cross}, tagging the protons and anti-protons in the forward proton detectors~\cite{FPD}. The TOTEM collaboration measured the $pp$ elastic $d\sigma/dt$ cross section at the LHC at center-of-mass energies $\sqrt{s}$ of 2.76, 7, 8 and 13 TeV~\cite{totemdata} in a wide kinematical region in $|t|$. The fact that it is possible to run the LHC machine at different $\sqrt{s}$ was fundamental for the odderon discovery. The additional advantage of the LHC machine is that it is possible to run it with different values of $\beta^*$ allowing to perform the elastic $d\sigma/dt$ cross section measurement over a wide kinematical region in $|t|$ between $10^{-4}$ and 3.5 GeV$^2$. The TOTEM collaboration has also an excellent coverage in the forward region using two telescopes called $T_1$ and $T_2$ covering the rapidity domains of $3.1<|\eta|<4.7$ and $5.3<|\eta|<6.5$, respectively and allowing to veto on any event with additional emitted particles after collision in addition to the two protons. 

The TOTEM measurements at different $\sqrt{s}$ are shown in Fig.~\ref{fig1}, left, and show always the same pattern. As a function of $|t|$, $d\sigma/dt$ decreases, reaching a minimum called the dip, then increases, reaching a maximum called the bump, and then decreases again. The same behavior is not observed for $p \bar{p}$ elastic interactions, where the $d\sigma/dt$ cross section does not show any dip nor bump~\cite{d0cross}. This difference of behavior between $pp$ and $p\bar{p}$ elastic interactions can be due to the existence of the odderon.


The strategy to measure quantitatively the differences between $pp$ and $p\bar{p}$ elastic interactions is to extrapolate the pp measurements from TOTEM at 2.76, 7, 8 and 13 TeV down to the Tevatron $\sqrt{s}$ of 1.96 TeV in order to compare directly with the D0 measurement. We thus first identify eight characteristic points in $|t|$ and $d\sigma/dt$ related to the shape of $pp$ elastic $d\sigma/dt$, namely the bump, the dip,  $dip2$ and $bump2$ corresponding to the same cross section at the dip and the bump at higher and lower $|t|$, respectively, $mid1$ and $mid2$ corresponding to the middle in $d\sigma/dt$ between the bump and the dip, and $bump+5$ and $bump+10$ corresponding to $d\sigma/dt$ at low $|t|$ with 5 or 10 times the differences between the cross section at the bump and the dip, as shown in Fig.~\ref{fig1}, right.

\begin{figure*}
\centering
\includegraphics[width=0.5\textwidth]{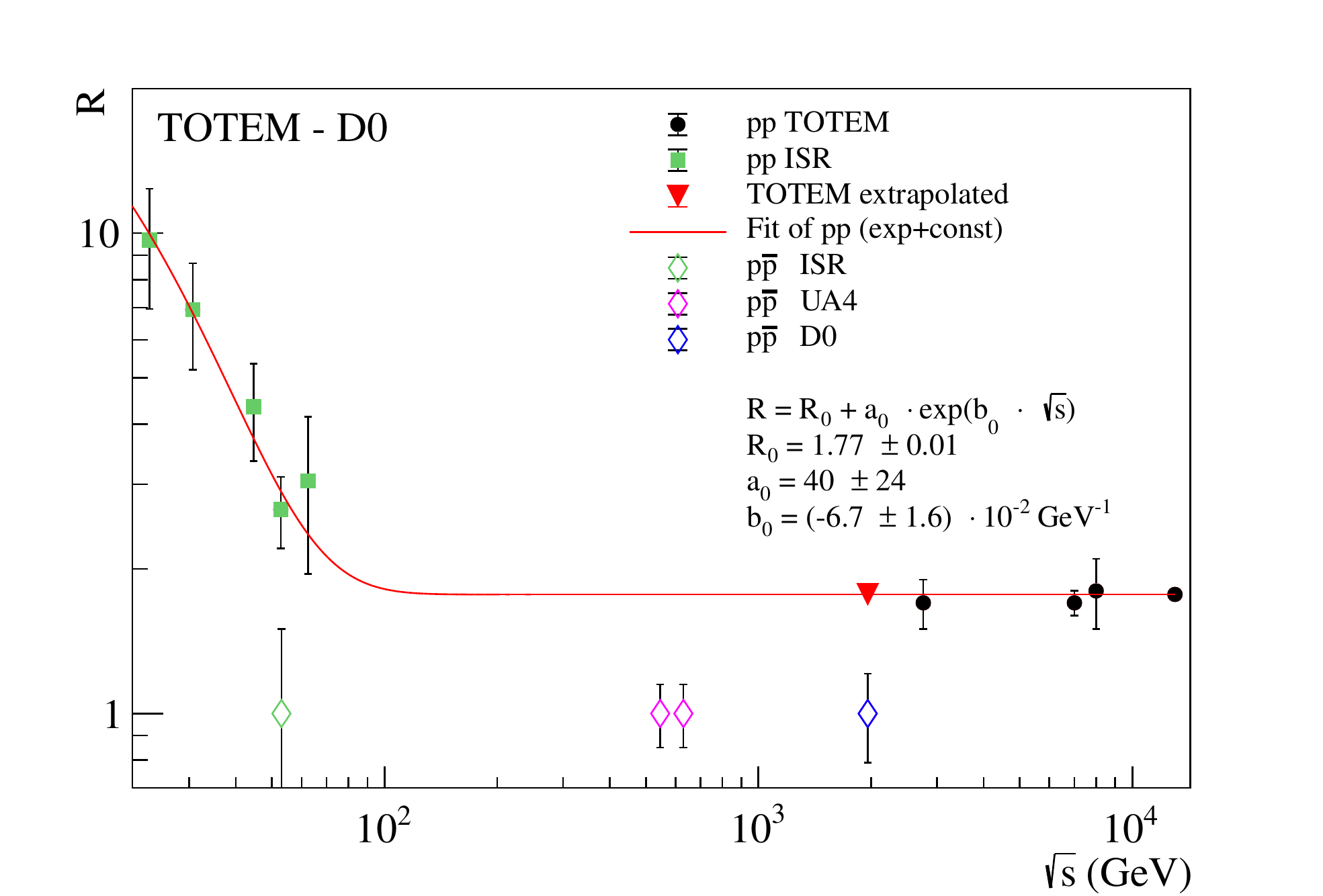}
\includegraphics[width=0.4\textwidth]{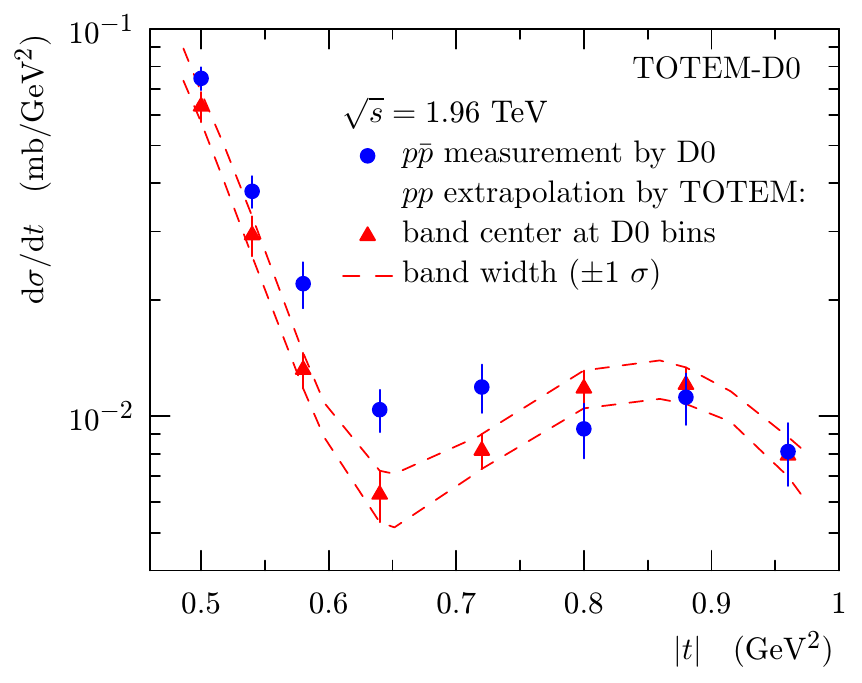}
\caption{Left: Bump over dip ratio for $pp$ and $p \bar{p}$ interactions as a function of the center-of-mass energies.
Right: Comparison at 1.96 TeV  between the elastic $d\sigma/dt$ cross section measurements for $p \bar{p}$ interactions at the Tevatron by the D0 experment and for $pp$ inetractions at the LHC, extrapolated from the TOTEM measurements at 2.76, 7, 8 and 13 TeV}
\label{fig2}
\end{figure*}

Measuring the $|t|$ and $d\sigma/dt$ values for each characteristic point as a function of $\sqrt{s}$ and fitting the $\sqrt{s}$ dependence as $|t| = a \log (\sqrt{s}{\rm [TeV]}) + b$ and 
$d\sigma/dt = c \sqrt{s}~{\rm [TeV]} + d$ allows predicting the values of the characteristic points at the Tevatron energy of 1.96 TeV. The last step is to fit the characteristic points at 1.96 TeV using a double exponential formula in order to predict the $pp$ elastic $d\sigma/dt$ cross sections in the same $|t|$ bins as for the D0 measurement for $p\bar{p}$ elastic interactions~\cite{ourpaper}. The differences in normalization are taken into account by
adjusting TOTEM and D0 data sets to have the same cross sections at the optical point
$d\sigma/dt(t = 0)$.

\subsection{The odderon discovery}
We first measured the ratio of the elastic cross sections at the dip and at the bump for $pp$ and $p \bar{p}$ interactions as shown in Fig.~\ref{fig2}, left. 
The bump over dip ratio in $pp$ elastic collisions decreases as a function of $\sqrt{s}$ up to $\sim$ 100 GeV~\cite{ISR} and is flat above that energy.
The $p \bar{p}$ ratios are constant at 1.00 (1.00$\pm$0.21 for D0) given the fact that no  bump nor dip is observed within uncertainties. It leads to a difference by more than 3$\sigma$ between $pp$ and $p \bar{p}$ elastic data, assuming a flat behavior above $\sqrt{s} = 100$ GeV. 

A more detailed comparison between the elastic D0 cross section measurement and the extrapolation of the TOTEM data down to the Tevatron $\sqrt{s}$ is shown in Fig.~\Ref{fig2}, right, in the dip and bump region. 
The $\chi^2$ test of the compatibility between the $pp$ and $p\bar{p}$ data with six degrees of freedom yields the $p$-value of 0.00061, corresponding to a significance of 3.4$\sigma$~\cite{ourpaper}. 

The combination with the independent evidence of the 
odderon found by the TOTEM Collaboration using  $\rho$~\cite{rho} and total cross section measurements at very low $t$ in a 
completely different kinematical domain corresponding to the Coulomb-nuclear interference region leads to a  
significance ranging from 5.3 to  5.7$\sigma$ depending on the model.
Models without colorless $C$-odd gluonic compound are thus excluded.

It is worth noticing that many additional studies in diffraction can be performed at the LHC and we can quote as examples the production of jets, $\gamma+$jets or $W/Z$ in single diffraction and double pomeron exchanges that will allow to constrain further the pomeron structure~\cite{cyrille,annabelle} or the gap between jet measurements that allow to probe Balitsky Fadin Kuraev Lipatov resummation effects~\cite{jetgapjet}.


\section{Anomalous coupling studies using intact protons at the LHC }

The same roman pot detectors can be used to look for beyond standard model physics at the LHC at low $\beta^*$ which is the standard high luminosity configuration at the LHC. The diffractive mass acceptance of the roman pot detectors to measure both intact protons in the final state starts at about 450 GeV and goes up to 2.4 TeV. In this section, we will study the exclusive production of diphotons, W and Z bosons, $\gamma+Z$, $t \bar{t}$,... by detecting the intact protons in the final state.

\subsection{Exclusive production of diphotons and sensitivity to axion-like particles}

\begin{figure*}
\centering
\includegraphics[width=0.4\textwidth]{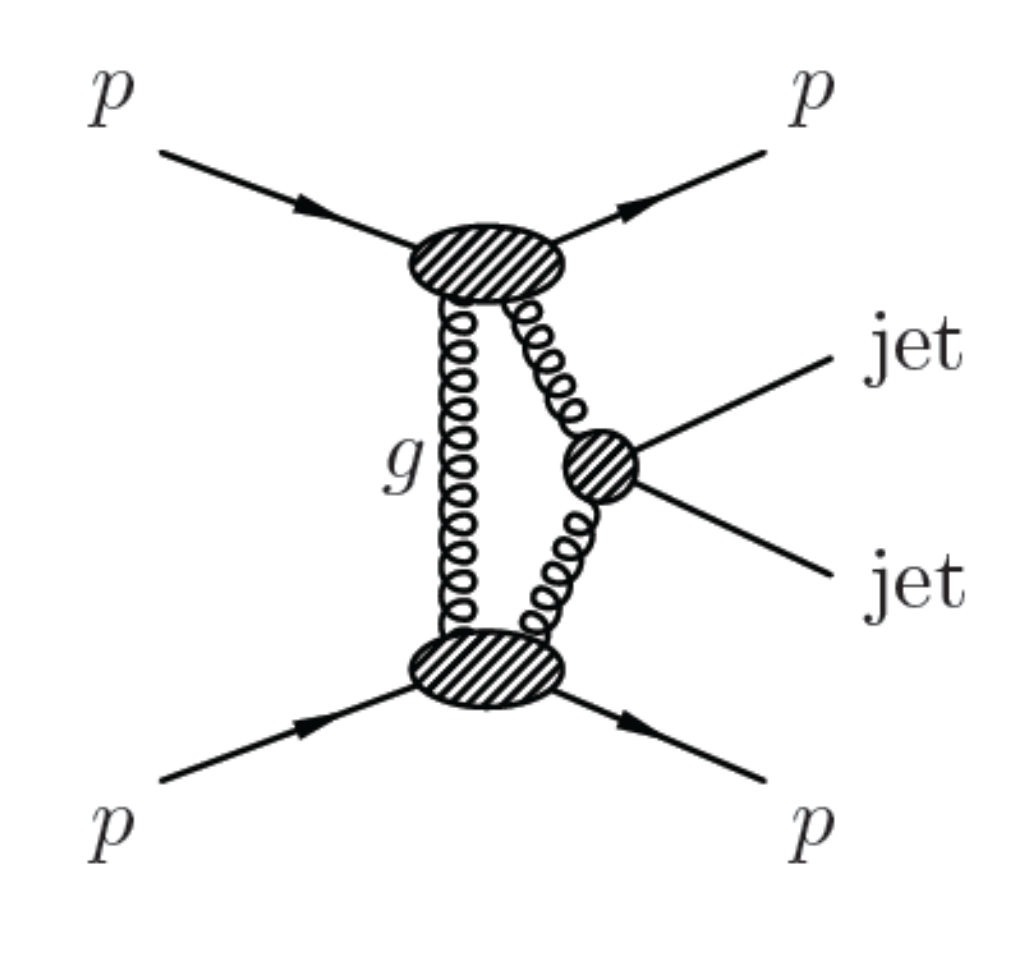}
\includegraphics[width=0.4\textwidth]{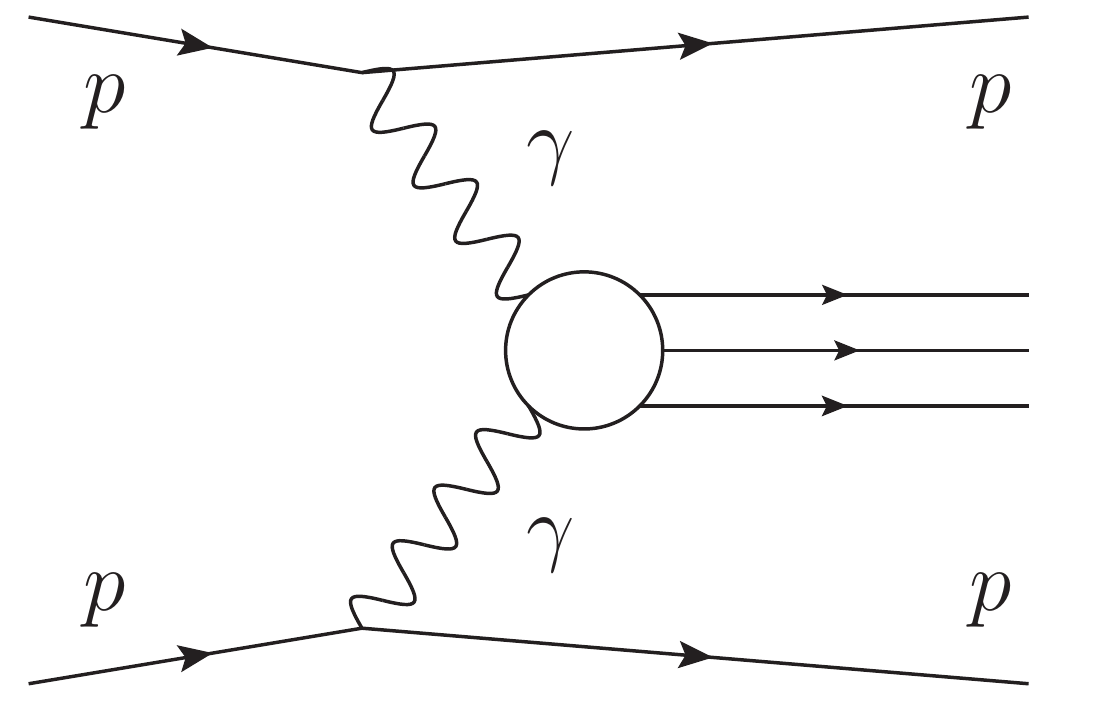}
\caption{Exclusive diphoton production via QCD and QED induced processes.}
\label{fig3b}
\end{figure*}

In Fig.~\ref{fig3b}, we display the two diagrams that lead to the production of diphotons and two intact protons in the final state. The left and the right ones correspond respectively to QCD and QED interactions. At high diphoton masses (above 450 GeV where the acceptance of the forward detectors starts), the QED one dominates by several orders of magnitude compared to the QCD one. It means that the observation in CMS and TOTEM of two  photons and two tagged protons is due to a photon-induced process.

The motivation to look for $\gamma \gamma \gamma \gamma$ $\zeta_1$ quartic coupling is twofolds. $\gamma \gamma \gamma \gamma$ couplings can be modified in a model
independent way by loops of heavy charge particles leading to 
$\zeta_1 = \alpha_{em}^2 Q^4 m^{-4} N c_{1,s}$ 
where the coupling depends only on $Q^4 m^{-4}$ (the charge and mass of the particle in the loop) and on spin, $c_{1,s}$. $\zeta_1$ can also be modified by neutral particles appearing as a resonance at tree level (extensions of the SM including scalar, pseudo-scalar, and spin-2 resonances that couple to the photon) $\zeta_1 =
(f_s m)^{-2} d_{1,s}$ where $f_s$ is the $\gamma \gamma X$ 
coupling of the new particle to the photon, and $d_{1,s}$ depends on the spin of the particle.  This
leads to typical values of $\zeta_1$ of the order of 10$^{-14}$-10$^{-13}$ GeV$^{-4}$.

\begin{figure*}
\centering
\includegraphics[width=0.9\textwidth]{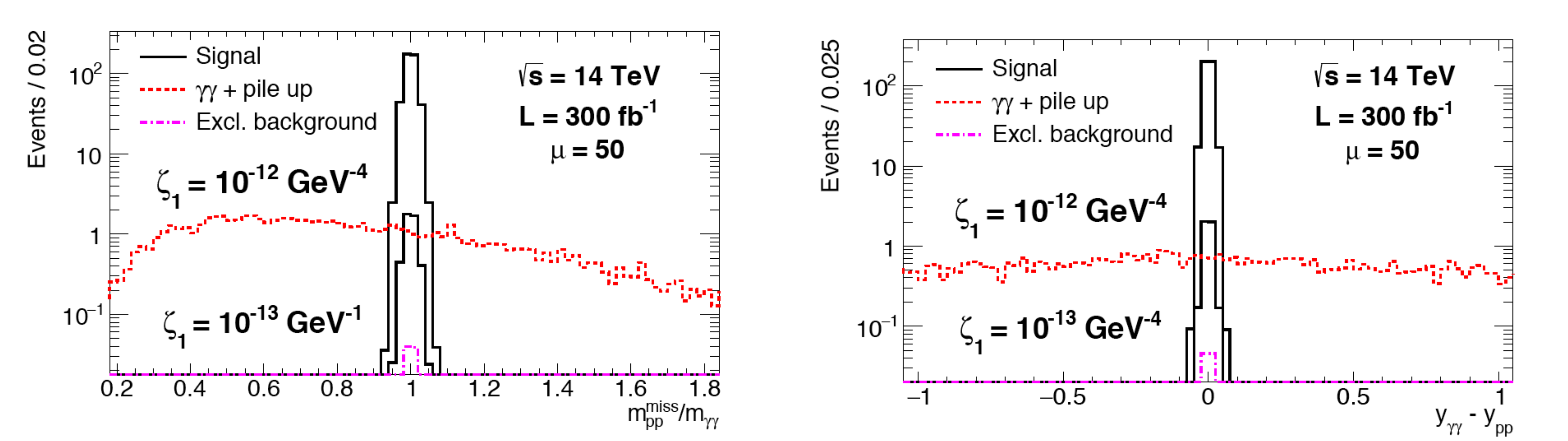}
\caption{Diffractive mass ratio and difference in rapidity using either the diphoton information from CMS or the proton taggings for signal in black and pile up background in red.}
\label{fig4}
\end{figure*}

The signature of exclusive diphoton events is very clean, two photons and two protons and nothing else, which leads to a powerful method to reject background. The main background to be considered is due to pile up where the two photon originate from one $pp$ inteaction and the two intact protons from additional pile up. Matching the mass and rapidity of the diphoton and proton information leads to a negligible background for 300 fb$^{-1}$ as shown in Fig.~\ref{fig4}~\cite{diphoton}. The CMS collaboration searched for exclusive diphoton production requesting back-to-back photons, at high mass ($m_{\gamma \gamma}>350$ GeV), and matching in rapidity and mass between diphoton and proton information. No signal was found and the first limits on quartic photon anomalous couplings at high mass were extracted, $|\zeta_1|<2.9~10^{-13}$ GeV$^{-4}$, $|\zeta_2|<6.~10^{-13}$ GeV$^{-4}$ with about 10 fb$^{-1}$. The sensitivities were updated using 102.7 fb$^{-1}$ with the following limits $|\zeta_1|<7.3~10^{-14}$ GeV$^{-4}$, $|\zeta_2|<1.5~10^{-13}$ GeV$^{-4}$~\cite{diphotoncms}.

\begin{figure*}
\centering
\includegraphics[width=0.7\textwidth]{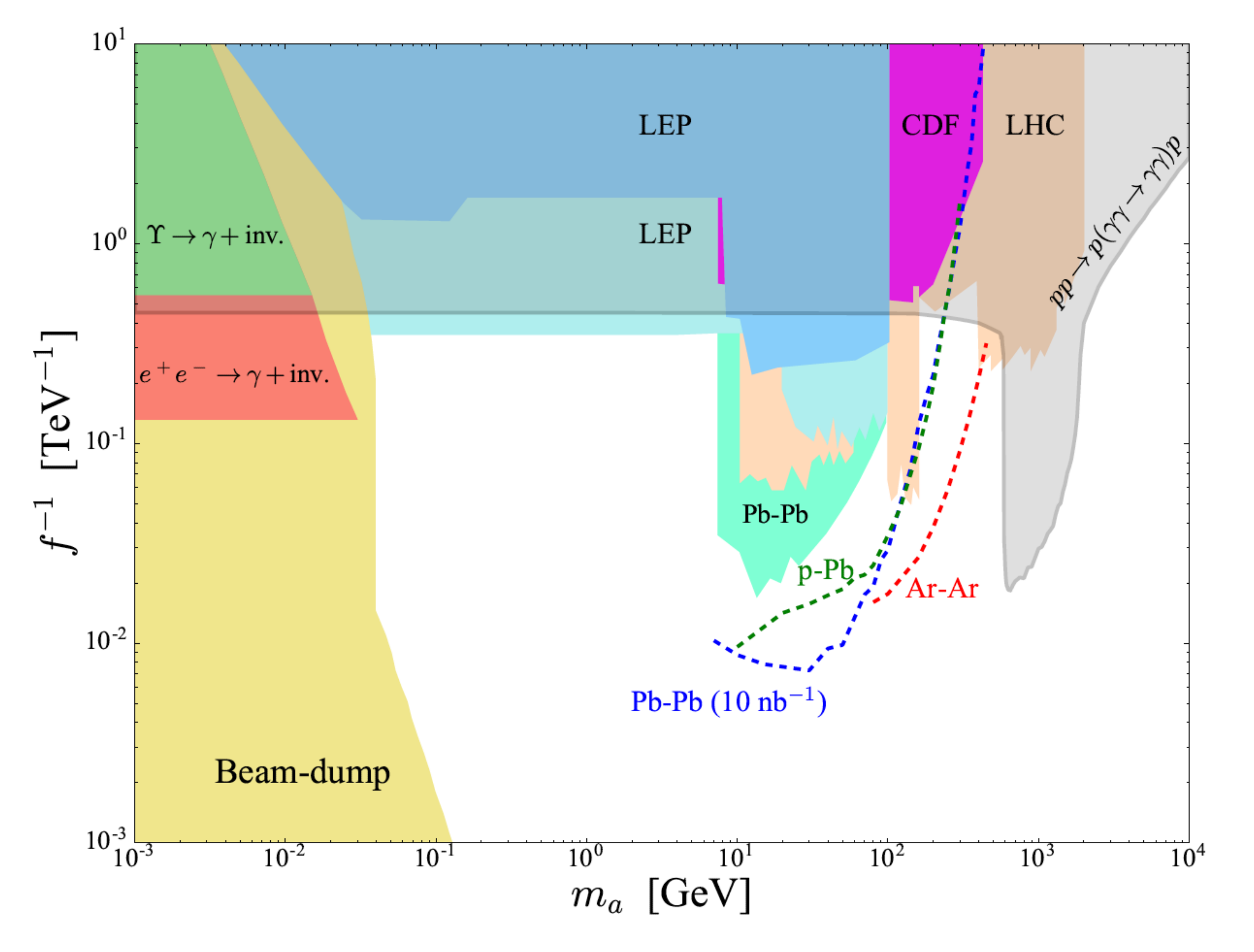}
\caption{Sensitivity plot in the coupling versus mass plane for ALP production.}
\label{fig5}
\end{figure*}

These results can be reinterpreted in terms of production of axion-like particles (ALP) that decay into two photons. The sensitivities for 300 fb$^{-1}$ in the $f^{-1}$ coupling versus axion mass $m_a$ plane are shown in Fig.~\ref{fig5}~\cite{alp}. It is worth noticing that the sensitivities increased by about two orders of magnitude at high ALP masses using this method and that a domain at very high mass is covered in addition with respect to more standard methods at the LHC. It is also complementary to the measurement of exclusive diphoton production in $pPb$, $PbPb$, $ArAr$ as shown in Fig.~\ref{fig5}. For heavy ion runs, there is no pile up, and the intact ions are not measured, and the rapidity gap method is used~\cite{alp}.

\subsection{Exclusive $WW$, $ZZ$, $\gamma Z$ and $t \bar{t}$ productions}

The strategy is to look for the SM exclusive production of $WW$ and $ZZ$ bosons in the lepton channel decay of the $W$ and $Z$ bosons since it leads to the cleanest signal and lowest backgrounds~\cite{exclww} leading to a possible observation of 50 signal events for 300 fb$^{-1}$  for 2 background events. The production of $WW$ and $ZZ$ exclusive events vis anomalous coupling appears at higher $WW$ and $ZZ$ and leads to a low cross section with respect to SM for many values of anomalous couplings. The search requires the highest branching ratios for the $W$ and $Z$ bosons, and thus their full hadronic decays. The CMS collaboration requested for existence the measurement of two ``fat" jets (of radius 0.8) corresponding to the $W$ decays, with a $p_T>200$ GeV, and a dijet mass 1126$<m_{jj}<$2500 GeV. The  jets are required to be back-to-back ($|1-\phi_{jj}/\pi|<0.01$).  As usual, the signal region is defined by the correlation between central $WW$ system and the proton information. The limits on quartic anomalous couplings were found to be $a_0^W/\Lambda^2 < 4.3~10^{-6}$ GeV$^{-2}$, $a_C^W/\Lambda^2 < 1.6~10^{-5}$ GeV$^{-2}$, $a_0^Z/\Lambda^2 < 0.9~10^{-5}$ GeV$^{-2}$, $a_C^Z/\Lambda^2 < 4.~10^{-5}$ GeV$^{-2}$ with 52.9 fb$^{-1}$~\cite{exclww}.

The same method applies to the search for $\gamma \gamma \gamma Z$ anomalous couplings when the $Z$ boson decays either into leptons or hadrons. The sensitivity improves by about three orders of magnitude using this method compared to other methods at the LHC looking for the $Z$ boson decay into three photons~\cite{exclgammaz}.

Additional searches include the search for $\gamma \gamma t \bar{t}$ anomalous couplings where one can look for the semi-lectonic and leptonic decays of the top and anti-top quarks. 
The CMS collaboration put a limit on the exclusive $t\bar{t}$ production cross section of $\sigma^{excl.}_{t \bar{t}} < 0.6$ pb  using 29.4 fb$^{-1}$ of data. This search can benefit heavily for the measurements of proton time-of-flights to ensure that the protons originate from the same vertex as the decay products of $t$ and $\bar{t}$. The typical sensitivity using 300 fb$^{-1}$ of data and similar selection (high $t \bar{t}$ mass, matching between $pp$ and $t \bar{t}$ information), as well as timing measurements, is of the order of  6 to 7 10$^{-11}$ GeV$^{-4}$~\cite{exclttbar}.

\subsection{Timing detectors and applications}
As we mentioned, it is useful to measure precisely the proton time-of-flight to reduce the pile up background with a precision of 15 to 20 ps. A possiblity is to use ultra-fast Silicon detectors or LGADs. In addition, We developed  these detectors and their readout electronics with two possible applications in mind, namely the measurement of cosmic ray radiation in space in collaboration with NASA and the measurement of doses in flash beam cancer therapy. UFSDs are typically fast detectors (the signal duration can be a few ns) and the readout system is based on the fast sampling technique, taking for instance 256 points on the rising part of the signal. 

\begin{figure*}
\centering
\includegraphics[width=0.45\textwidth]{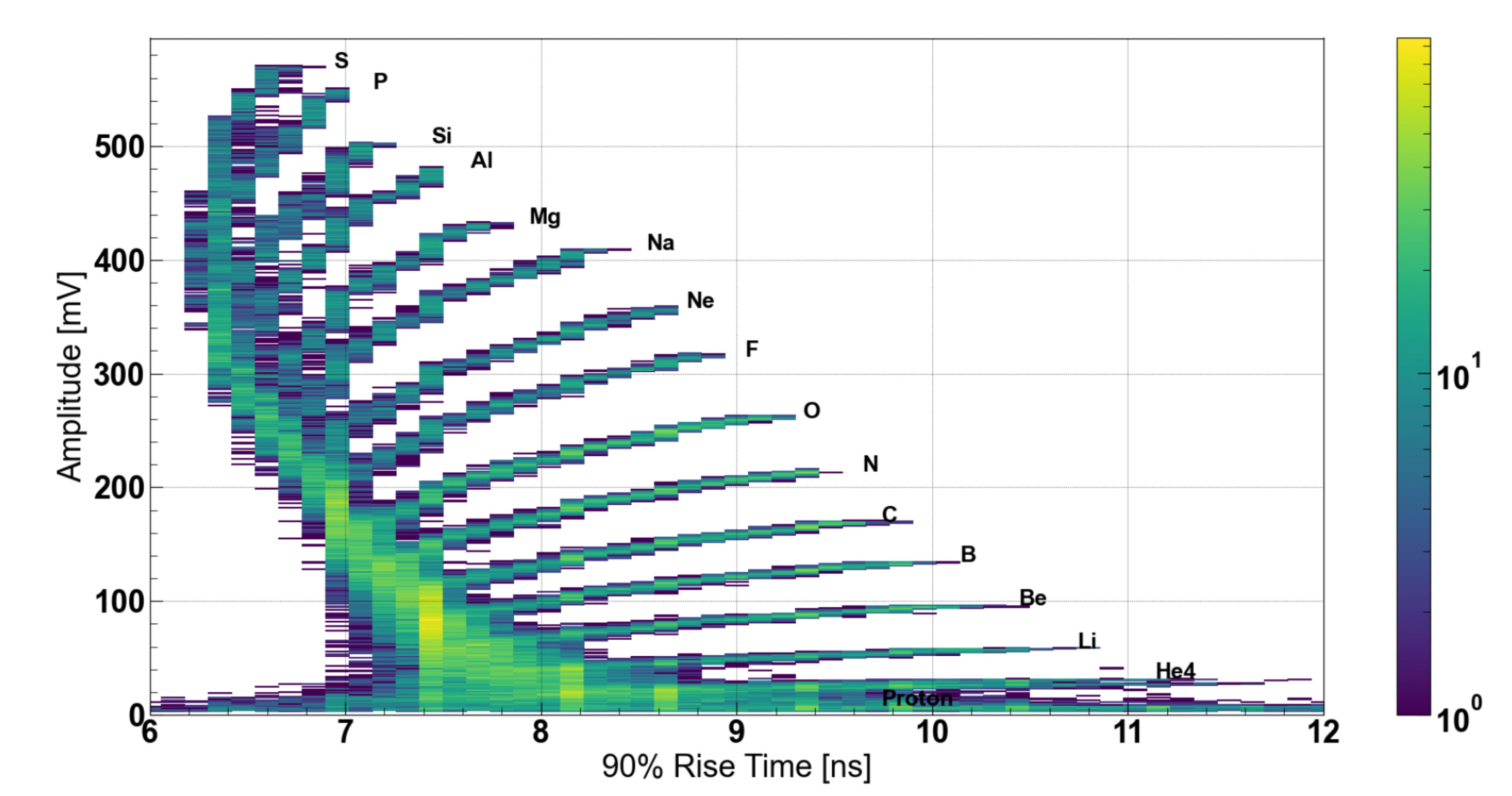}
\includegraphics[width=0.45\textwidth]{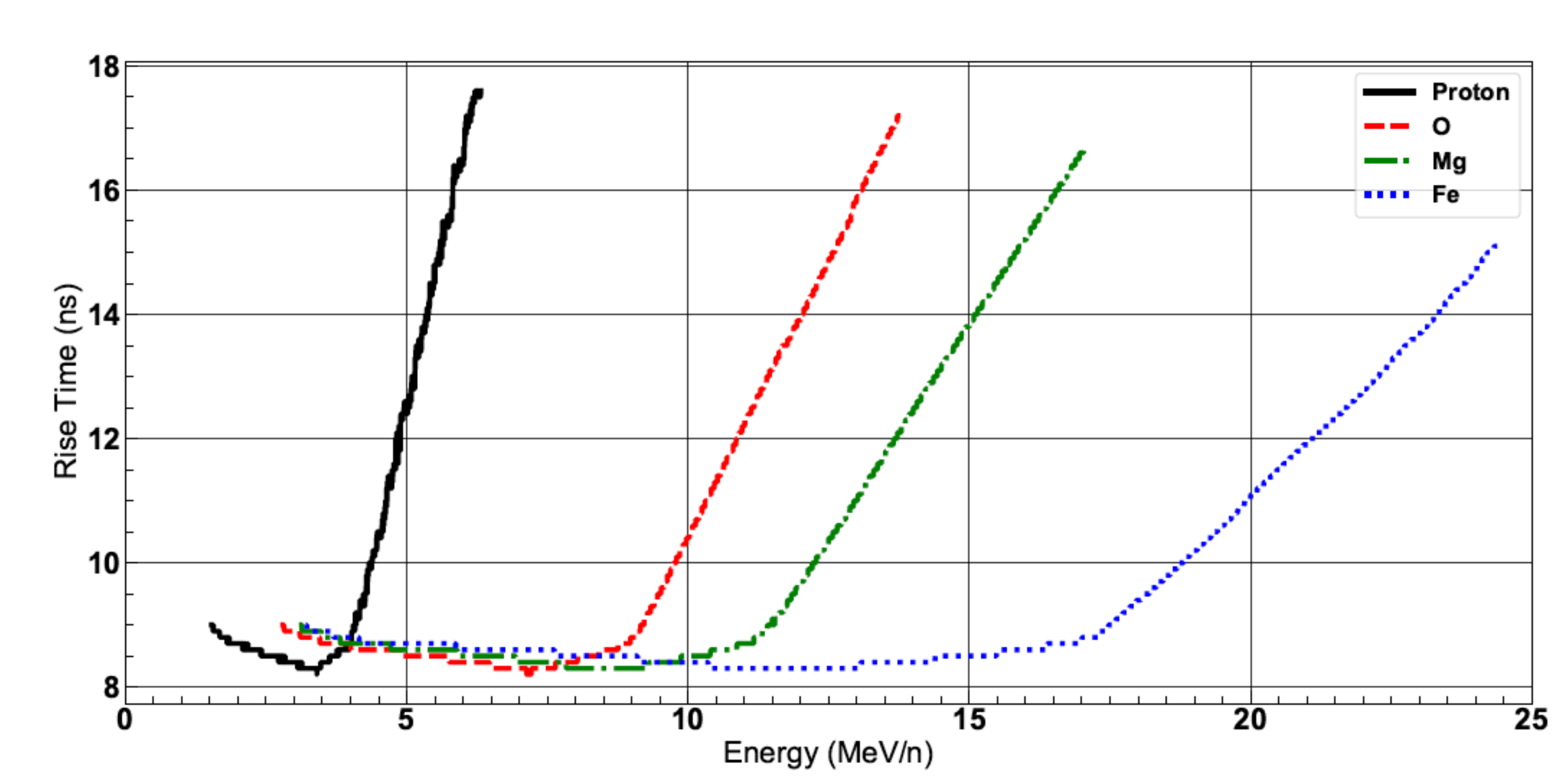}
\caption{Left: Signal amplitude versus 90\% of rise time showing the possibility to identify the type of particle. Right: Rise time vs energy showing the possibility of measuring the particle energy once it is identified.}
\label{fig6b}
\end{figure*}

Using this method, we can determine the duration and amplitude of the signal of a stopping particle in a given layer of UFSD that allows to determine the type of particle in cosmic rays (p, He, Fe, Pb...) as well as their energy~\cite{nasa}. This is illustrated in Fig.~\ref{fig6b} where we show the correlation between the signal amplitude and the rise time for different kinds of particles (left) and between the rise time and the energy of the particle (right) that allows to determine the particle energy once its type is known. This is the first time that the fast sampling technique will be used in space.  

\begin{figure*}
\centering
\includegraphics[width=0.45\textwidth]{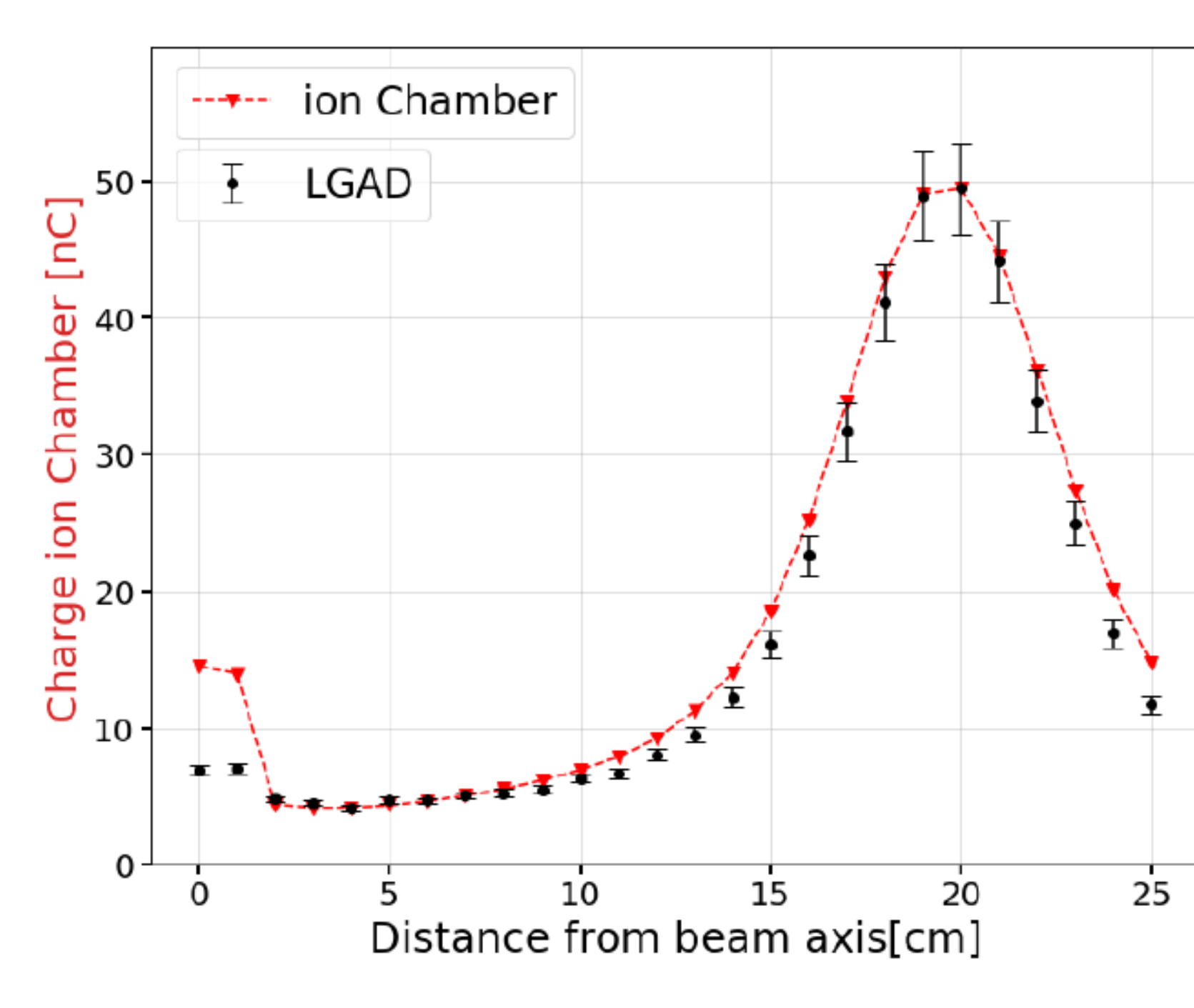}
\includegraphics[width=0.45\textwidth]{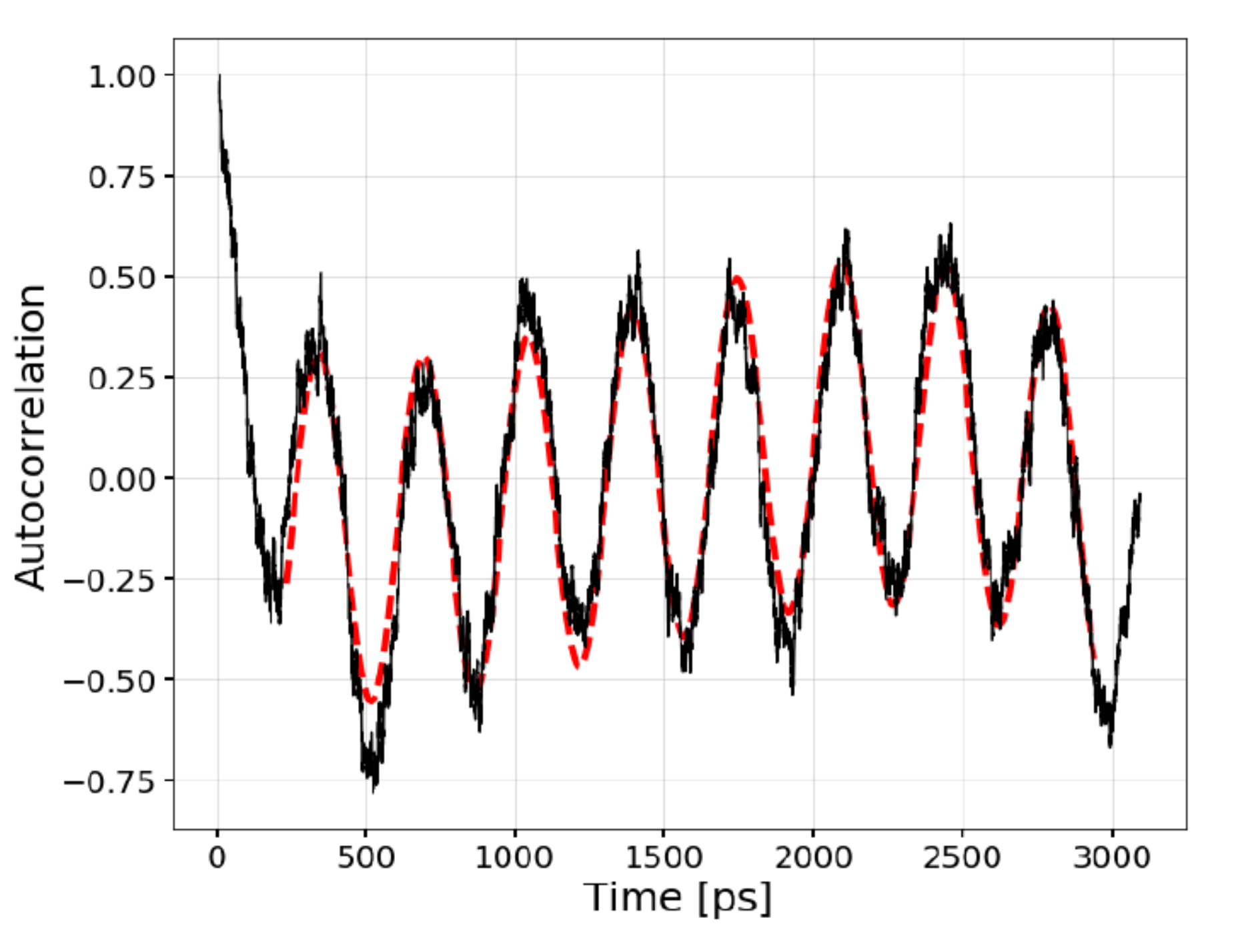}
\caption{Left: Comparison between the measurement of LGAD and ion chamber. Right: Beam structure analysis in Dublin hospital using the UFSD detector.}
\label{fig6}
\end{figure*}

The same method can be used in medical physics. The idea is to use the fast aspect of the UFSD signal to count the number of protons/electrons, and thus determine the delivered dose to patients, in flash beam therapy used in cancer treatments. In order to show this possibility, we put our UFSD in the beam used for therapy at St Luke hospital in Dublin and we saw for the first time the beam structure as shown in Fig~\ref{fig6} while the correlation with the usual dose measurement using ion chambers is very good.

\section{Conclusion}

In this report, we first discussed the odderon discovery by the D0 and TOTEM experiment. We then described the improvement on sensitivities to beyond standard model physics via the observation of quartic anomalous couplings by 2 or 3 orders of magnitude compared to usual LHC measurements using intact protons in the final state. We finally described briefly the ultra fast Silicon detectors that we use to measure cosmic rays or doses in flash beam therapy as an application to the detectors originally developed for high energy and nuclear physics.

\end{multicols}

\medline

\begin{multicols}{2}

\end{multicols}
\end{document}